\documentclass[12pt,fleqn]{article}
\usepackage{amsmath}
\usepackage{slashed}
\usepackage{url}
\usepackage{bm}
\usepackage{graphicx}
\usepackage[utf8]{inputenc}
\usepackage[english]{babel}
\usepackage{amsmath}
\usepackage{amsfonts}
\usepackage{amssymb}
\usepackage{graphicx}
\usepackage{float}
\usepackage{hyperref}

\begin{document}

\title{Non-minimally coupled scalar field and scaling symmetry in a cosmological background}
\author{Malik Almatwi\footnote{malik.almatwi@gmail.com}\,\, }

\date{Department of Theoretical Physics, Faculty of Science, University of Mazandaran, 47416-95447, Babolsar, Iran}
\maketitle

\tableofcontents
\section{Abstract}
We study scaling symmetry in a class of non-minimally coupled scalar field in a background of Friedmann-Robertson-Walker (FRW) spacetime. We use a non-minimally coupling $R L^{(\varphi)}$. We find the corresponding conserved charge of that symmetry and see its role in cosmology, and search for its possible breaking down and its outcomes. A suitable potential $V(\varphi)=\varphi^2/2$ of scalar field is adopted which is necessary to get a scaling symmetric Lagrangian of the system including scalar field, non-minimally coupling to Ricci scalar $R L^{(\varphi)}$ and dark matter dust. We study evolution of the scalar field in the phase space of the model and explore the stability of the obtained critical point. In this manner we derive a relation that relates the cosmological constant and gravitational constant via a unique identity which reflects the scaling symmetry breaking in the space $(a, \varphi)$. And relate the cosmological constant to the vacuum expectation value of the potential energy of $\varphi$. Finally we study the stability of that vacuum expectation value.

\vspace{\baselineskip}
{\bf Keywords}: Scalar Field Cosmology, Non-minimal Gravitational Coupling; Noether Symmetry, Cosmic Speed Up.

\section{Introducing a Scaling Symmetric Lagrangian in Non-Minimally Coupled Scalar Field in FRW spacetime}
A scaling symmetry in the space $(a, \varphi)$ is a kind of unification of real scalar field $\varphi(t)$(represents dark energy) with the universal scale factor $a(t)$ (represents spatial homogenous FRW metric), that is, existence one of them implies existence the other, by that the energy and geometry can be accompanied in one identity, i.e, we can put them in one field such as $\Phi=(a, \varphi)$ on a manifold without needing introducing any geometry and then we can establish a Lagrangian in terms of $\Phi$ that respect the corresponding symmetry(but in this paper we do not do that). By that we can explain the relation between energy and geometry in more general concept( i.e, symmetry concept), by which the gravity is explained by concept of scaling symmetry group. Note that there is dissimilarity with symmetry of general relativity in which the transformations do not change $\varphi$ regarding it as a scalar field, unlike the scaling symmetry, in which the scalar field changes with changing the metric.

\vspace{\baselineskip}
In this paper, we use a non-minimally coupled scalar field to gravity by the term $R L^{(\varphi)}$, we use FRW metric and find a global scaling symmetry Lagrangian in the space $(a, \varphi)$, whose symmetry breaking yields the usual Lagrangian of the non-minimally coupled scalar field. The scaling symmetry Lagrangian in the space $(a, \varphi)$ implies existence a globally conserved quantity (charge) which can be used for global classification of the cosmological solutions, i.e, two solutions with unequal charges can not be related to each other by any coordinates transformation. We treat the role of the charge in the solutions of $\varphi$ and we show that by the universal positively accelerated expansion (increasing the scale factor $a$ exponentially) the field $\varphi$ is always exponentially decreasing until reaching a critical point in $\dot \varphi=0$ with $\varphi=\varphi_0 \ne0$, in which the global scaling symmetry breaks and the universal expansion is approximately in a constant rate $H=H_0$. 

\vspace{\baselineskip}
The evidence of existence of symmetry breaking is seen by violating conservation of the corresponding charge; $dQ/dt\ne0$. We will find that symmetry breaking occurs in the critical point $\dot \varphi=0$, $\varphi=\varphi_0 \ne0$. The existence of a non-vanishing constant value $\varphi_0$ at that critical point $\dot \varphi=0$ is needed for satisfying the constraint equation $\delta S /\delta N=0$. We find that the critical point $\dot \varphi=0$, $\varphi=\varphi_0 \ne0$ is unique and stable (there are no other critical points). As a result, we can relate the cosmological constant and gravitational constant to a same identity, which is scaling symmetry breaking in the space $(a, \varphi)$. And relate the cosmological constant to the vacuum expectation value of the potential energy of $\varphi$. If we thank that the vacuum expectation value of $\varphi$ does not depend on any metric and it is a quantum phenomena, we obtain universal constant vacuum energy(cosmological constant). By that we relate the cosmological constant to a quantum phenomena, but here the field is specified by a scaling symmetry Lagrangian (\ref{eq:2}).

\vspace{\baselineskip}
The Lagrangian of the gravity plus the scalar field can be written in the background of the spatially flat FRW metric $ds^{2}=-N(t)dt^{2}+a^{2}(t)(dx^{2}+dy^{2}+dz^{2})$ as a one-point Lagrangian up to boundary terms as~\cite{Remmen2013}
\begin{equation}
\begin{split}
 \sqrt g L(N, a,\dot a,\varphi ,\dot \varphi ) &= \frac{1}{{16\pi G}}\sqrt { - g} R + \sqrt { - g} L^{(\varphi )}  \\ 
 & =  - 3m_{pl}^2 N a^3 \left( \frac{{\dot a^2 }}{N^2 a^2}  \right)+N a^3 \left( {\frac{1}{2}\frac{{\dot \varphi ^2 }}{N^2} - V\left( \varphi  \right)} \right) + {\text{boundary terms}}\, .
\end{split}
\end{equation}

Where $N(t)$ is lapse function and $a(t)$ is cosmic scale factor. We note that $a$ and $\varphi$ are dynamical variables, while $N(t)$ is a non-dynamical variable, it just represents the symmetry in direction of time, therefore we choose $N(t)=1$ after deriving the equations of motion.\\

In this paper we study the scalar fields coupled non-minimally to gravity as $RL^{(\varphi)}$, where $R$ is Ricci scalar and $L^{(\varphi)}$ is Lagrangian of scalar field $\varphi$. Let us introduce a non-minimally coupled scalar field to gravity by
\begin{equation}
\begin{split}
 \sqrt { - g} L(N, a,\dot a,\varphi ,\dot \varphi ) =  - 3m_{pl}^2 a\frac{{\dot a^2 }}{N} &+ a^3 \left( {\frac{1}{2}\frac{{\dot \varphi ^2 }}{N} - NV\left( \varphi  \right)} \right) \\ 
&  + 3kNa\frac{{\dot a^2 }}{{N^2 }}\left( {\frac{1}{2}\frac{{\dot \varphi ^2 }}{N^2} - V\left( \varphi  \right)} \right)\,,  
\end{split}
\end{equation}
in which we have used some constant $k>0$ for satisfying the units. Here the non-minimal interaction(third term) of the scalar field with gravity is represented by product of the scalar $Na\left( {\dot a /N} \right)^2$ with the Lagrangian $\dot \varphi ^2 /2N^2 - V\left( \varphi  \right)$ of scalar field. We note that there is no problem with that coupling since both $Na\left( {\dot a} /N\right)^2$ and $L^{(\varphi)}$ are scalars, therefore their product is also scalar and preserves all of 
their symmetries. 

\vspace{\baselineskip}
For more general case, we add dust matter (visible and dark) density $\rho _m \left( a \right)=\rho _0^{(m)} /a^3$, where $\rho _0^{(m)}$ is a constant as matter density at some scale factor $a_0=1$, and a cosmological constant $\Lambda$ which will just become as a result of global scaling symmetry breaking of the Lagrangian (\ref{eq:2}), while the dust matter does not effect on the results of that global symmetry and its breaking, and setting $\rho _0^{(m)}=0$ is possible, but we add it just to notify that dark matter dust can exist in the phase of global scaling symmetry. By adopting the scalar field  potential of the form $V\left( \varphi  \right) =\varphi^2 /2$ which is needed to obtain a scaling symmetric Lagrangian (\ref{eq:2}), we get
\begin{equation}
\begin{split}
 \sqrt { - g} L(N, a,\dot a,\varphi ,\dot \varphi ,\rho _m ) =&  - 3m_{pl}^2 a\frac{{\dot a^2 }}{N} + a^3 \left( {\frac{1}{2}\frac{{\dot \varphi ^2 }}{N} - \frac{1}{2}N\varphi ^2 } \right) \\ 
&+ 3ka\frac{{\dot a^2 }}{{N^2 }}\left( {\frac{1}{2}\frac{{\dot \varphi ^2 }}{N} - \frac{1}{2}N\varphi ^2 } \right) - Na^3 \rho _m \left( a \right) - Na^3 \Lambda \,.  
\end{split}
\end{equation}
Thus we obtain a one point Lagrangian of gravity + NMC term + scalar field + matter density in the minimum super-space $(a,\varphi)$ of the model. If we let the variables to be measured in units of Planck mass, we set $m_{pl}=1$ to get
\begin{equation}\label{eq:1}
\begin{split}
 \sqrt { - g} L(N, a,\dot a,\varphi ,\dot \varphi ,\rho _m ) =&  - 3a\frac{{\dot a^2 }}{N} + \frac{{a^3 }}{2}\left( {\frac{{\dot \varphi ^2 }}{N} - N\varphi ^2 } \right) \\ 
&  + \frac{{3k}}{2}\frac{{a\dot a^2 }}{{N^2 }}\left( {\frac{{\dot \varphi ^2 }}{N} - N\varphi ^2 } \right) - N\rho _0^{(m)}  - Na^3 \Lambda \,.  
\end{split}
\end{equation}
We can let this Lagrangian be produced from another Lagrangian which has a global scaling symmetry in the space $(a,\varphi ) $, like the Lagrangian
\begin{equation}\label{eq:2}
\sqrt { - g} L\left( {N, a,\dot a,\varphi ,\dot \varphi } \right) = \frac{{a^3 }}{2}\left( {\frac{{\dot \varphi ^2 }}{N} - N\varphi ^2 } \right) + \frac{{3k}}{2}\frac{{a\dot a^2 }}{{N^2 }}\left( {\frac{{\dot \varphi ^2 }}{N} - N\varphi ^2 } \right) - N\rho _0^{(m)} \,.
\end{equation}
This Lagrangian includes a scalar field Lagrangian with interaction with gravity in addition to a dark matter density term $ \rho_m (a)=\rho _0^{(m)} / a^3$. Thus we have a global scaling symmetry in the space of dynamical variables $a$ and $\varphi$ represented by the transformations
\begin{equation}\label{eq:18}
a \to e^{2\alpha } a,\quad \textrm{and} \quad \varphi  \to e^{ - 3\alpha } \varphi \, ,
\end{equation}
for an arbitrary real constant parameter $\alpha$ which can be either positive or negative. Thus the Lagrangian (\ref{eq:2}) has global scaling symmetry $L\left( {e^{2\alpha } a ,e^{ - 3\alpha } \varphi  } \right) =L\left( {a,\varphi } \right) $ in the space $(a,\varphi ) $, but this symmetry is broken when there is a non-vanishing ground state value of $\varphi^2$, like $\left\langle \Omega  \right|\varphi ^2 \left| \Omega  \right\rangle  = \varphi _0^2  \ne 0$, for a ground state wave function $ \left| \Omega  \right\rangle$ of the Lagrangian (\ref{eq:2}). This implies a replacing $ \varphi ^2$ with $ \varphi ^2+ \varphi ^2_0$ nearby the minimum energy state $\left| \Omega  \right\rangle$ in the Lagrangian (\ref{eq:2}) to get
\begin{equation}
\begin{split}
 L&\left( {N,a,\dot a,\varphi ,\dot \varphi } \right) \\ 
  &= \frac{{a^3 }}{2}\left( {\frac{{\dot \varphi ^2 }}{N} - N\varphi ^2  - N\varphi _0^2 } \right) + \frac{{3k}}{2}\frac{{a\dot a^2 }}{{N^2 }}\left( {\frac{{\dot \varphi ^2 }}{N} - N\varphi ^2  - N\varphi _0^2 } \right) - N\rho _0^{(m)}  \\ 
 & =  -\varphi _0^2 \frac{{3k}}{2} \frac{{a\dot a^2 }}{N} + \frac{{a^3 }}{2}\left( {\frac{{\dot \varphi ^2 }}{N} - N\varphi ^2 } \right) + \frac{{3k}}{2}\frac{{a\dot a^2 }}{{N^2 }}\left( {\frac{{\dot \varphi ^2 }}{N} - N\varphi ^2 } \right) \\ 
&  - N\rho _0^{(m)}  - \frac{{\varphi _0^2 }}{2}Na^3 \,.  
\end{split}
\end{equation}
If we choose $k$ and $\varphi _0$ to get 
\begin{equation}\label{eq:20}
k \varphi _0^2/2=1,\quad and \quad \varphi _0^2/2=\Lambda, \quad for \quad N(t)=1 \, ,
\end{equation}
thus we get the Lagrangian (\ref{eq:1}) with scaling symmetry breaking. In terms of Planck mass, we get $k \varphi _0^2/2=m^2_{pl}$ so $k \Lambda=m^2_{pl}$ which unifies the gravitational constant with the cosmological constant via the scaling symmetry breaking. While $\varphi _0^2/2=\Lambda$ relates cosmological constant $\Lambda$ to  vacuum energy $ \varphi _0^2/2$ of scalar field. Actually the equation $k \varphi _0^2/2=1$ ensures that $k>0$, otherwise we will not get the usual general relativity of FRW metic as a result of scaling symmetry breaking in the space $\left(a, \varphi \right)$. As we will see that symmetry breaking occurs in the critical point $\dot \varphi=0$, $\varphi=\varphi_0 \ne0$ and this critical point is stable and unique.

\vspace{\baselineskip}
Now we derive the conserved charge and the equations of motions of the Lagrangian (\ref{eq:2}). Since $L\left( {e^{2\alpha } a ,e^{ - 3\alpha } \varphi  } \right) =L\left( {a,\varphi } \right) $, the action $S= \int L dt$ is also invariant. Therefore,
\[
\delta _\alpha  S = \int {dt\delta _\alpha  L}  = \int {dt\left( {L\left( {e^{2\alpha } a,e^{ - 3\alpha } \varphi } \right) - L\left( {a,\varphi } \right)} \right)}  = 0 \, .
\]
If we use an infinitesimal transformation $\alpha\ll 1$, we get
\[
\delta _\alpha  a = e^{2\alpha } a - a \approx \left( {1 + 2\alpha } \right)a - a = 2\alpha a \, ,
\]
and
\[
\delta _\alpha  \varphi  = e^{ - 3\alpha } \varphi  - \varphi  \approx \left( {1 - 3\alpha } \right)\varphi  - \varphi  =  - 3\alpha \varphi \, .
\]
Using these results in the following relation
\begin{equation}\label{eq:3}
\begin{split}
 \delta _\alpha  S &= \int {dt\delta _\alpha  L}  \\
 & =  - \int {dt\left(  \frac{d}{{dt}}\frac{{\partial L}}{{\partial \dot a}}  -\frac{{\partial L}}{{\partial a}}\right)}  - \int {dt\left(  \frac{d}{{dt}}\frac{{\partial L}}{{\partial \dot \varphi }}- \frac{{\partial L}}{{\partial \varphi }} \right)}  \\
&  + \int {dt\frac{d}{{dt}}\left( {\frac{{\partial L}}{{\partial \dot a}}\delta _\alpha  a + \frac{{\partial L}}{{\partial \dot \varphi }}\delta _\alpha  \varphi } \right)}  = 0\,  ,
\end{split}
\end{equation}
with regarding the equations of motions, we obtain a conserved charge as
\[
Q = \frac{{\partial L}}{{\partial \dot a}}\left( {2a} \right) + \frac{{\partial L}}{{\partial \dot \varphi }}\left( { - 3\varphi } \right),\quad \frac{dQ}{{dt}} = 0\, .
\]
Therefore we get(With using the gauge $N(t)=1$)
\begin{equation}\label{eq:4}
\begin{split}
 Q &= 3ka\dot a\left( {\dot \varphi ^2  - \varphi ^2 } \right)\left( {2a} \right) + \left( {a^3  + 3ka\dot a^2 } \right)\dot \varphi \left( { - 3\varphi } \right) \\
 & = 6ka^3 H\left( {\dot \varphi ^2  - \varphi ^2 } \right) - 3a^3 \left( {1 + 3kH^2 } \right)\frac{d}{{dt}}\left( {\frac{{\varphi ^2 }}{2}} \right) \\
  &= 12ka^3 Hp - \frac{3}{2}a^3 \left( {1 + 3kH^2 } \right)\left( {\dot \rho  - \dot p} \right)=constant \, .
\end{split}
\end{equation}
In which we have used $H\equiv\dot a /a$, the energy density $\rho\equiv\dot \varphi ^2 /2 + \varphi ^2/2$ and the momentum density (pressure density) $p\equiv\dot \varphi ^2 /2 - \varphi ^2/2$ of $\varphi$. 

\vspace{\baselineskip}
We note that for a solution like $H=H_0$, $\dot \rho  = \dot p=0$ and $\varphi=\varphi_0=constant \ne0$ (that is, $\dot \varphi =0$), we have
\begin{equation}\label{eq:5}
\left. {\frac{{dQ}}{{dt}}} \right|_c  =- 12ka^3 \left( {3H_0^2  + \dot H_0} \right)\rho _0  \ne 0\, ,
\end{equation}
where the non-vanishing value in the right side comes from slow-rolling condition $ \dot H_0 \approx 0$. Thus in this case, the scaling symmetry of the Lagrangian (\ref{eq:2}) breaks and by that we get the Lagrangian (\ref{eq:1}). Actually we will find that the point $H=H_0=constant$, $\varphi=\varphi_0=constant \ne0$ is a stable critical point for the dynamical system of the Lagrangian (\ref{eq:2}) and it is a unique critical point.

\vspace{\baselineskip}
The equation of motions of $a$ from the Lagrangian (\ref{eq:2}), $\delta S /\delta a=0$(With using the gauge $N(t)=1$), is
\[
\frac{d}{{dt}}\left( {\frac{{\partial L}}{{\partial \dot a}}} \right) - \frac{{\partial L}}{{\partial a}} = 0 \, ,
\]
which yields
\begin{equation}
\begin{split}
& 3k\dot a\dot a\left( {\dot \varphi ^2  - \varphi ^2 } \right) + 3ka\ddot a\left( {\dot \varphi ^2  - \varphi ^2 } \right) + 6ka\dot a\frac{d}{{dt}}\left( {\frac{{\dot \varphi ^2 }}{2} - \frac{{\varphi ^2 }}{2}} \right) \\
&  - \frac{{3a^2 }}{2}\left( {\dot \varphi ^2  - \varphi ^2 } \right) - \frac{{3k}}{2}\dot a^2 \left( {\dot \varphi ^2  - \varphi ^2 } \right) = 0 \, ,
\end{split}
\end{equation}
and by using $H=\dot a /a$, $\ddot a /a=\dot H+H^2$ and the momentum density $p=\dot \varphi ^2 /2 - \varphi ^2/2$, the last equation becomes
\[
\left( {6k\dot H + 9kH^2  - 3} \right)p + 6kH\frac{{dp}}{{dt}} = 0\, .
\]
Using a dimensionless time parameter defined as $\eta  = \ln \left( {a/a_0 } \right)$ which regards the scale factor $a$ as a cosmological time, we have $d/dt=H d/d\eta$. The last equation becomes
\[
\left( {6kHH' + 9kH^2  - 3} \right)p + 6kH^2 p' = 0\, ,
\]
or
\begin{equation}\label{eq:5}
\left( {h' + 3h - 3} \right)p + 2hp' = 0\, ,
\end{equation}
where the prime indicates the derivative with respect to the dimensionless time $\eta$, and we used $h=3kH^2$ as a dimensionless function. 

\vspace{\baselineskip}
The equation of motion of $\varphi$ from the Lagrangian (\ref{eq:2}), $\delta S /\delta \varphi=0$(With using the gauge $N(t)=1$), is
\[
\left( {3a^2 \dot a + 3k\dot a\dot a^2  + 6ka\dot a\ddot a} \right)\dot \varphi  + \left( {a^3  + 3ka\dot a^2 } \right)\ddot \varphi + a^3 \varphi  + 3ka\dot a^2 \varphi = 0\, .
\]
Following the same steps as for $a$, we obtain
\begin{equation}\label{eq:6}
\left( {h' + 3h + 3} \right)\left( \rho+p \right) + \left( { 1+h} \right)\rho' = 0\, ,
\end{equation}
where we used the energy density $\rho=\dot \varphi ^2 /2 + \varphi ^2/2$ of $\varphi$.

\vspace{\baselineskip}
We note that the previous equations of $\rho$ and $p$ include $h'$, so we need to omit it, by that the two equations (\ref{eq:5}) and (\ref{eq:6}) give one equation.

\vspace{\baselineskip}
The constraint equation of the Lagrangian (\ref{eq:2}), $\delta S /\delta N=0$, implies
\[
\frac{{a^3 }}{2}\left( { - \frac{{\dot \varphi ^2 }}{{N^2 }} - \varphi ^2 } \right) - 3k\frac{{a\dot a^2 }}{{N^3 }}\left( {\frac{{\dot \varphi ^2 }}{N} - N\varphi ^2 } \right) + \frac{{3k}}{2}\frac{{a\dot a^2 }}{{N^2 }}\left( { - \frac{{\dot \varphi ^2 }}{{N^2 }} - \varphi ^2 } \right) - \rho _0^{(m)}  = 0\, .
\]
Using the gauge $N(t)=1$, we obtain
\[
\frac{{a^3 }}{2}\left( {\dot \varphi ^2  + \varphi ^2 } \right) + \frac{{9k}}{2}a\dot a^2 \dot \varphi ^2  - \frac{{3k}}{2}a\dot a^2 \varphi ^2  + \rho _0^{(m)}  = 0\, ,
\]
or
\[
2a^3 \rho  + 3a^3 h\left( {\rho  + p} \right) - a^3 h\left( {\rho  - p} \right) + 2\rho _0^{(m)}  = 0\, .
\]
Therefore we obtain the energy constraint equation
\begin{equation}\label{eq:7}
\rho  + h\left( {\rho  + 2p} \right) + \frac{{\rho _0^{(m)} }}{{a^3 }} = 0\, .
\end{equation}
We note that the energy constraint (\ref{eq:7}) does not include any critical energy(such as $3H^2$). But it imposes some conditions, since $\rho=\dot \varphi ^2 /2 + \varphi ^2/2$, $h=3kH^2$ and $\rho _0^{(m)}\ne 0$ are always positive, we have $\rho  + 2p<0$ and it must be always satisfied. Therefore the pressure $p=\dot \varphi ^2 /2 - \varphi ^2/2$ must be always negative, $p<0$, and does not vanish.  However negative pressure is needed for getting universal expansion.\\

This means that the potential energy $ \varphi ^2/2$ is always larger than the kinetic energy $\dot \varphi ^2/2$, therefore there is no possibility to increase the kinetic energy and vanishing the potential energy, while the opposite is possible, that is increasing in potential energy while decreasing in kinetic energy until it vanishes. Thus the solution $\dot \varphi =0$, $ \varphi =\varphi_0=constant\ne0$ is possible. \\

Since $\rho  + 2p<0$ and $p<0$, we obtain $\rho  + 3p<0$ which according to Friedmann equations implies an universal accelerated expansion.

\vspace{\baselineskip}
We also note that the case $\rho=0$ ($\varphi=0$) does not exist since it implies $p=0$. So, we have $0+ {{\rho _0^{(m)} }}/{{a^3 }} = 0$ which is not satisfied unless $\rho _0^{(m)}=0$. Therefore the acceptable minimum energy is $\rho_0\ne 0$ (for $\dot \varphi =0$) and this value corresponds to the vacuum expectation value of $\varphi ^2$, as discussed just after equation (\ref{eq:18}).

\vspace{\baselineskip}
The energy constraint equation (\ref{eq:7}) does not give $h$ as a function only of $\rho$ and $p$, in addition it includes $a(t)$. Therefore we need to omit $h'$ from the two equations (\ref{eq:5}) and (\ref{eq:6}) to get
\begin{equation}\label{eq:9}
\left( {1 + h} \right)p\rho ' - 2h\left( {\rho  + p} \right)p' + 6p\left( {\rho  + p} \right) = 0\, .
\end{equation}
The same equation we will obtain if we get $h'$ from the constraint equation (\ref{eq:7}) and use it in the equations (\ref{eq:5}) and (\ref{eq:6}). 

\vspace{\baselineskip}
In order to get another equation for $\rho'$ and $p'$, we omit $1/ a^3$ from the charge equation (\ref{eq:4}) and constraint equation (\ref{eq:7}). We obtain
\[
12kHp - \frac{3}{2}\left( {1 + h} \right)H\left( {\rho ' - p'} \right) + c\rho  + ch\left( {\rho  + 2p} \right) = 0,\quad \textrm{for} \quad c = \frac{Q}{{\rho _0^{(m)} }} \, ,
\]
which gives
\begin{equation}\label{eq:8}
\rho ' - p' = \frac{{8kp}}{{\left( {1 + h} \right)}} + \frac{{2c\rho }}{{3H\left( {1 + h} \right)}} + \frac{{2ch}}{{3H\left( {1 + h} \right)}}\left( {\rho  + 2p} \right) \, .
\end{equation}
Since both $H$ and $h>0$ can not vanish for any solution, there is no problem with $H\left( {1 + h} \right)$ in the denominator of the last equation. 

\vspace{\baselineskip}
By that we have two equations, (\ref{eq:9}) and (\ref{eq:8}), that include $\rho'$, $p'$, $\rho$, $p$ and $h=3kH^2$. From these equations, we obtain
\begin{equation}\label{eq:10}
\begin{split}
&\left[2h\left( {\rho  + p} \right)  - \left( {1 + h} \right)p   \right]\rho ' \\
&=   \left( {\rho  + p} \right)\left[ {\frac{{16khp}}{{\left( {1 + h} \right)}} + \frac{{4ch\rho }}{{3H\left( {1 + h} \right)}} + \frac{{4ch^2 }}{{3H\left( {1 + h} \right)}}\left( {\rho  + 2p} \right) + 6p} \right]\\
&= \left( {\rho  + p} \right)\left[ {{4ckH}\rho  + \frac{{16khp}}{{\left( {1 + h} \right)}} + \frac{{8ch^2 }}{{3H\left( {1 + h} \right)}}p + 6p} \right]\, ,
\end{split}
\end{equation}
and
\begin{equation}\label{eq:11}
\left[ 2h\left( {\rho  + p} \right)  - \left( {1 + h} \right)p  \right]p' =  8kp^2 + \frac{{2c}}{{3H}}p\rho  + 2ckHp\left( {\rho  + 2p} \right) + 6p\left( {\rho  + p} \right) \, .
\end{equation}
We have $p<0$, $h>0$ and $\rho  + p=\dot \varphi ^2 \ge 0$, therefore it is always $[2h\left( {\rho  + p} \right)  - \left( {1 + h} \right)p] >0$ and does not vanish. Thus there is no problem with multiplying $\rho '$ and $p'$ by $[2h\left( {\rho  + p} \right)  - \left( {1 + h} \right)p] $.

\section{Critical points and scaling symmetry breaking}

We note that the constraint equation (\ref{eq:7}) does not imply any critical energy (such as $3H^2$), so we do not need to divide $\rho$ and $p$ by any energy and since we set $m_{pl}=1$, the variables $\rho$, $p$, $H$, $a$ and $\eta=\ln(a)$ are dimensionless. Thus the critical points of the equations (\ref{eq:10}) and (\ref{eq:11}) can be obtained by finding the points of $\rho '=p'=0$, at a time $\eta_0=\ln(a_0)$, in the space $(\rho , p)$, where $H$ can be written in terms of these quantities. We note that the time $\eta_0=\ln(a_0)$ does not mean to stop universal expansion, but it is just point in the space $(\rho , p)$, and nearby that point the velocity $(\rho' (\eta) , p' (\eta))$ decreases till finish at the point $\eta_0=\ln(a_0)$. So this does not mean stop universal expansion, but it is just a moment of it(at $a_0=a(t_0)$). And since velocity $(\rho' , p')$ is infinitesimal in vicinity of the point $\rho '=p'=0$, thus the evolution of the system nearby that point needs largest times, so the time is most spent in vicinity of critical points $\rho '=p'=0$. Therefore the solutions near by the critical points characterizes the solutions of the system in good accepted approximation, i.e, solutions in $t=\pm \infty$ or at $t=t_0$. 

\vspace{\baselineskip}
We note that since the scale factor $a(t)$ is assumed always in increasing, so indeed the energy density $\rho$ of the scalar field is in decreasing till reaching a smallest possible value at $\rho '=p'=0$($\eta_0=\ln(a_0)$). We denote $(\rho_0 , p_0)$ as a critical point ($\rho '=p'=0$) and this critical point belongs to a trajectory in the space $(\rho , p)$ where this trajectory is parameterized by the time parameter $\eta=\ln(a)$. Therefore, the critical point $(\rho_0 , p_0)$ is determined by the time $\eta_0=\ln(a_0)$ on that trajectories. Thus, for each critical point ($\rho'=p'=0$), we have the quantities of $\rho_0$, $p_0$, $H_0$ and $\eta_0=\ln(a_0)$. As we will show there is only one critical point associated with the scaling symmetry breaking of the Lagrangian (\ref{eq:2}).

\vspace{\baselineskip}
The condition $\rho'=0$(equation (\ref{eq:10})) gives the following two equations,
\begin{equation}\label{eq:15}
4ckH\rho  + \frac{{16khp}}{{\left( {1 + h} \right)}} + \frac{{8ch^2 }}{{3H\left( {1 + h} \right)}}p + 6p = 0 \, ,
\end{equation}
and
\begin{equation}\label{eq:16}
 \rho  + p= 0 \, .
\end{equation}
While the condition $p'=0$(equation (\ref{eq:11})) gives only one equation (with $p\ne0$) as
\begin{equation}\label{eq:12}
8kp^2  + \frac{{2c}}{{3H}}p\rho  + 2ckHp\left( {\rho  + 2p} \right) + 6p\left( {\rho  + p} \right) = 0\, .
\end{equation}
While the energy constraint (\ref{eq:7}) implies(at $\rho'=p'=0$)
\[
\left. {h'} \right|_c \left( {\rho _0  + 2p_0 } \right) - \frac{{3\rho _0^{(m)} }}{{a_0^4 }}\left. {a'} \right|_c  = 0\, ,
\]
and by using
\[
a' = \frac{{\partial a}}{{\partial \eta }} = \frac{{\partial a}}{{\partial \ln \left( a \right)}} = a\frac{{\partial a}}{{\partial a}} = a\, ,
\]
we get the equation
\begin{equation}\label{eq:a1}
\left. {h'} \right|_c \left( {\rho _0  + 2p_0 } \right) - \frac{{3\rho _0^{(m)} }}{{a_0^3 }} = 0\, ,
\end{equation}
which determines $h'$ at the critical point $\rho'=p'=0$. Note that $\left. {h'} \right|_c=0$ is satisfied only when $\rho _0^{(m)}=0$(so getting de Sitter solution). However if we assume that ${\rho _0^{(m)} }/{a_0^3 } $ is small enough, which implies $\left. {h'} \right|_c \approx 0$ (so $\dot H \approx 0$), we obtain solutions close to de Sitter solution(we will find that in slow-rolling condition).

\vspace{\baselineskip}
In fact, the two equations (\ref{eq:15}) and (\ref{eq:12}) disagree, therefore the critical points are given only by the two equations (\ref{eq:16}) and (\ref{eq:12}). We can see this  disagreement if we multiply the equation (\ref{eq:15}) by $3H\left( {1 + h} \right)/2\ne 0$, to get
\begin{equation}\label{eq:13}
2c\rho h + 2ch^2 \left( {\rho  + 2p} \right) + 24khpH + 9pH\left( {1 + h} \right) = 0\, .
\end{equation}
While, multiplying equation (\ref{eq:12}) by $3Hh\ne 0$ and dividing it by $p \ne 0$ with using $h=3kH^2$, we find
\begin{equation}\label{eq:14}
24kHhp + 2c\rho h + 2ch^2 \left( {\rho  + 2p} \right) + 18Hh\left( {\rho  + p} \right) = 0\, .
\end{equation}
Now subtracting equation (\ref{eq:13}) from equation (\ref{eq:14}), we obtain
\begin{equation}\label{eq:17}
-9pH\left( {1 + h} \right) + 18Hh\left( {\rho  + p} \right) = 0\, .
\end{equation}
But as we saw, the pressure $p$ in this setup is always negative and non-vanishing, $p<0$ (which comes from the conditions $\rho\ne 0$ and $(\rho  + 2p)<0$), and also $H>0$ does not vanish, while $(\rho  + p)\ge 0$, thus the last equation is sum of two positive terms and one of them does not vanish, so their sum also does not vanish. Therefore the last equation can not be satisfied as required. So, the two equations (\ref{eq:15}) and (\ref{eq:12}) disagree and the critical points $\rho '=p'=0$ are described only by two equations (\ref{eq:16}) and (\ref{eq:12}).

\vspace{\baselineskip}
From the equation (\ref{eq:16}), we get $p_0  =  - \rho _0  < 0$, using it in equation (\ref{eq:12}), we get
\[
3ckH^2_0  + 12kH_0 - c = 0\, .
\]
Its positive solution is
\[
H_0  = \frac{{ - 2}}{c} + \sqrt {\frac{4}{{c^2 }} + \frac{1}{{3k}}}  = \frac{{ - 2\rho _0^{(m)} }}{Q} + \sqrt {\left( {\frac{{2\rho _0^{(m)} }}{Q}} \right)^2  + \frac{1}{{3k}}}\, .
\]
From the equation of the charge (\ref{eq:4}), we get
\[
Q = 12ka^3 Hp - \frac{3}{2}a^3 \left( {1 + 3kH^2 } \right)\left( {\dot \rho  - \dot p} \right) =  - 12ka_0^3 H_0 \rho _0\, .
\]
But the quantities $a_0$, $H_0$, and $ \rho _0$ are all positive, therefore $Q$ is negative. Thus we replace $Q\to -Q$ to get a positive quantity for our forthcoming purpose. In this manner we obtain the expansion rate at the critical point as follows
\[
H_0  = \frac{{2\rho _0^{(m)} }}{Q} + \sqrt {\left( {\frac{{2\rho _0^{(m)} }}{Q}} \right)^2  + \frac{1}{{3k}}} >0\, .
\]
We note that for $\rho _0^{(m)} \ll \rho _0$, this expansion rate approximates to $H_0  = 1/ \sqrt {3k}$ which agrees with slow rolling solution.

\vspace{\baselineskip}
Now we show that the conservation of the charge (\ref{eq:4}) is broken at this critical point. We have
\begin{equation}
\begin{split}
 \left. {Q'} \right|_c & = \left. {\frac{{dQ}}{{d\eta }}} \right|_c  =  - 12ka^3 \left. {\left( {3H + H'} \right)} \right|_c \rho _0  =  - \frac{{12ka^3 }}{{3kH_0 }}\left. {\left( {9kH^2  + 3kHH'} \right)} \right|_c \rho _0  \\
 & =  - \frac{{12ka^3 }}{{3kH_0 }}\left. {\left( {3h + \frac{1}{2}h'} \right)} \right|_c \rho _0  \, ,
\end{split}
\end{equation}
where we have used $\rho''=p''=0$ because $\rho'\sim (\rho-\rho_0)$ and $p'\sim (p-p_0)$, so $\rho''\sim (\rho'-\rho_0)$ and $p''\sim (p'-p_0)$, therefore $\rho''=p''=0$ at the critical point $(\rho_0, p_0 )$.\\

From the equations (\ref{eq:7}) and (\ref{eq:a1}), we obtain
\begin{equation}\label{eq:19}
\left. h \right|_c  = 1 + \frac{{\rho _0^{(m)} }}{{\rho _0 a^3_0 }}, \quad \textrm{and} \quad \left. {h'} \right|_c  =  - \frac{{3\rho _0^{(m)} }}{{\rho _0 a^3_0 }}\, .
\end{equation}
Using these relations in $Q'$, we get
\[
\left. {Q'} \right|_c  =  - \frac{{12ka_0^3 }}{{3kH_0 }}\left( {3 + \frac{{3\rho _0^{(m)} }}{{2\rho _0 a_0^3 }}} \right)\rho _0  \ne 0\, .
\]
In this situation, the scaling symmetry of the Lagrangian (\ref{eq:2}) is broken at the critical point $\dot \varphi=0$, $\varphi(a_0)=\varphi_0\ne 0$, at time $\eta_0=\ln(a_0)$, thus we get the Lagrangian (\ref{eq:1}). We note that for $\rho _0^{(m)} \ll \rho _0$, we have $ \left. {h'} \right|_c  \approx0$ implying $H=H_0=constant$, which agrees with the  slow rolling solution and indicates  that nearby the critical point $\dot \varphi \approx0$, $\varphi_0 \ne 0$, the universal expansion rate becomes constant and we obtain a de Sitter solution.\\

We note that the quantities $\varphi_0$ and $H_0$ do not need to depend on $\eta_0=\ln(a_0)$, so only indeed the matter dust $\rho^{(m)} \sim 1/a^3$ will depend on $a_0$, thus we are free in choosing $a_0$ to get a suitable $\rho _0^{(m)}$ at point of scaling symmetry breaking.

\vspace{\baselineskip}
Now we show that the critical point $\dot \varphi=0$, $ \varphi=\varphi_0>0$ is stable. We find first order approximation of $\rho'$ and $p'$ nearby the critical point $(\rho_0, p_0 )$; $\rho_0 + p_0 =0$. Actually according to the equations (\ref{eq:19}), and with $\rho _0^{(m)} \ll \rho _0$, we can neglect perturbations on $h$ and so on $H$; $ \delta H \sim 1/a_0^3<<1$. 

\vspace{\baselineskip}
We have, the equations (\ref{eq:10}) and (\ref{eq:11}),
\begin{equation}
\begin{split}
&\left[2h\left( {\rho  + p} \right)  - \left( {1 + h} \right)p   \right]\rho ' \\
&= \left( {\rho  + p} \right)\left[ {{4ckH}\rho  + \frac{{16khp}}{{\left( {1 + h} \right)}} + \frac{{8ch^2 }}{{3H\left( {1 + h} \right)}}p + 6p} \right]\,,
\end{split}
\end{equation}
and
\begin{equation}
\left[ 2h\left( {\rho  + p} \right)  - \left( {1 + h} \right)p  \right]p' =  8kp^2 + \frac{{2c}}{{3H}}p\rho  + 2ckHp\left( {\rho  + 2p} \right) + 6p\left( {\rho  + p} \right) \,.
\end{equation}
Multiplying first equation by ${3H\left( {1 + h} \right)/2}$ and using $h=3k H^2$, we obtain
\begin{equation}
\begin{split}
 &\frac{{3H\left( {1 + h} \right)}}{2}\left[ {2h\left( {\rho  + p} \right) - \left( {1 + h} \right)p} \right]\rho ' \\ 
 & = \left( {\rho  + p} \right)\left[ {2ch\left( {1 + h} \right)\rho  + 24kHhp + 4ch^2 p + 9H\left( {1 + h} \right)p} \right] \\ 
 & = \left( {\rho  + p} \right)\left[ {2ch\rho  + 2ch^2 \rho  + 24kHhp + 4ch^2 p + 9H\left( {1 + h} \right)p} \right]\\
& = \left( {\rho  + p} \right)\left[ {24kHhp + 2ch\rho  + 2ch^2 \left( {\rho  + 2p} \right) + 9H\left( {1 + h} \right)p} \right]\, .
\end{split}
\end{equation}
Thus, nearby $\rho_0 + p_0 =0$ and by using the equation (\ref{eq:14}) (equation of $p'=0$), we get first order approximation
\begin{equation}
\begin{split}
\left( {1 + h_0 } \right)^2 \rho _0 \rho ' &= \left( {\Delta \rho  + \Delta p} \right)\left[ {-12h\left( {\rho_0  + p_0} \right)- 6\left( {1 + h_0} \right)\rho _0 } \right]\\
&\\
&\to \left( {\Delta \rho  + \Delta p} \right)\left[ {- 6\left( {1 + h_0} \right)\rho _0 } \right]\, ,
\end{split}
\end{equation}
so
\[
\left( {1 + h_0 } \right)\rho _0 \rho ' = \left( {\Delta \rho  + \Delta p} \right)\left( {- 6\rho _0 } \right)\, \Rightarrow \rho ' = \frac{{ - 6}}{{1 + h_0 }}\left( {\Delta \rho  + \Delta p} \right)\, ,
\]
for $\Delta \rho =\rho-\rho_0 \ll 1$ and $\Delta p =p-p_0 \ll 1$. Using this equation in the first order perturbation of equation (\ref{eq:8}), we get
\[
p' = \frac{{ - 2}}{{1 + h_0 }}\left[ {3\Delta \rho  + \left( {3 + 4k} \right)\Delta p} \right]\, .
\]
From last two equations, we obtain $(\lambda_1, \lambda_2)$ the eigenvalues of the velocities $(\rho ', p')$ nearby $(\rho_0, p_0 )$, we get
\[
\lambda _1  =  - \frac{1}{{1 + h_0 }}\left( {6 + 4k - 2\sqrt {4k^2  + 9} } \right)\approx  - \left( {3 + 2k - \sqrt {4k^2  + 9} } \right)\, ,
\]
and
\[
\lambda _2  =  - \frac{1}{{1 + h_0 }}\left( {6 + 4k + 2\sqrt {4k^2  + 9} } \right)\approx  - \left( {3 + 2k+ \sqrt {4k^2  + 9} } \right)\, .
\]
Since $k>0$(regarding equation (\ref{eq:20})), it is always $(6 + 4k - 2\sqrt {4k^2  + 9}) >0$, therefore both $\lambda _1 $ and $\lambda _2$ are negative, thus the critical point $(\rho_0, p_0 )$; $\rho_0 + p_0 =0$, $\rho_0>0$ is stable. Therefore the global scaling symmetry breaking is inevitable matter, and it is global critical point since it depends on vacuum energy of the scalar field $\varphi(t)$, which can be related to quantum phenomena(i.e, quantization, bosonic fields,...).\\

\section{Slow Rolling Solutions}
According to the equation (\ref{eq:a1}), in all critical points, we have $\left. {h'} \right|_c \approx 0$ when ${\rho _0^{(m)} }/{a_0^3 } $ is small enough(such ${\rho _0^{(m)} }/{a_0^3 } <<1$). This condition yields to the slow-rolling conditions $\left| {\ddot \varphi} \right|\ll\left| {\varphi} \right|$ and $\left| {\dot \varphi} \right|\ll\left| {\varphi} \right|$ which take place nearby the critical point $\dot \varphi=0$, $\varphi(a_0)=\varphi_0\ne 0$ of the scaling symmetry Lagrangian (equation (\ref{eq:2})), that yields to solutions close to de Sitter solution(universal expansion with constant rate $H=constant$). The necessity of slow-rolling solutions is in their obtaining the behaviour of all variables nearby the critical point $\dot \varphi=0$, $\varphi(a_0)=\varphi_0\ne 0$ and before the scaling symmetry breaking. As usual, we get the equation of expansion rate $H$ from the energy constraint equation (equation (\ref{eq:7})). We obtain
\begin{equation}
\begin{split}
 h = 3kH^2  &= \frac{{ - \rho  - \frac{{\rho _0^{(m)} }}{{a^3 }}}}{{\rho  + 2p}} = \frac{{ - \dot \varphi ^2  - \varphi ^2  - \frac{{\rho _0^{(m)} }}{{a^3 }}}}{{\dot \varphi ^2  + \varphi ^2  + 2\dot \varphi ^2  - 2\varphi ^2 }} \\
  &= \frac{{ - \dot \varphi ^2  - \varphi ^2  - \frac{{\rho _0^{(m)} }}{{a^3 }}}}{{3\dot \varphi ^2  - \varphi ^2 }} \Rightarrow \frac{{ - \varphi ^2  - \frac{{\rho _0^{(m)} }}{{a^3 }}}}{{ - \varphi ^2 }} = 1 + \frac{{\rho _0^{(m)} }}{{\varphi ^2 a^3 }} \approx 1 + \frac{{\rho _0^{(m)} }}{{\varphi_0 ^2 a^3 }}\, ,
\end{split}
\end{equation}
where we have used the slow rolling condition $\left| {\dot \varphi} \right|\ll\left| {\varphi} \right|$. If we impose a condition as ${{\rho _0^{(m)} }}/{{\varphi_0 ^2 a^3 }}\ll1$ which takes place at large scale factor values $a\gg1$ and with $\rho _0^{(m)}\ll \varphi _0^2/2$ for which the universe is dominated by the ground state energy of $\varphi$(vacuum energy) which has the role of cosmological constant, however, we identified the energy $ \varphi _0^2/2$ with cosmological constant, formulas (\ref{eq:20}). We obtain $3kH^2 \approx 1$, therefore we get approximately constant expansion rate $H_0=1/\sqrt{3k}$. Note that phase occurs at late times $a\gg1$ of universal expansion. Using $h'=0$, $h=1$ in the equation (\ref{eq:6}), we obtain
\[
6\dot \varphi ^2  + 2\left( {\dot \varphi \ddot \varphi  + \varphi \dot \varphi } \right) = 0 \,\,\, \Rightarrow\,\,\, 6\dot \varphi  + 2\left( {\ddot \varphi  + \varphi } \right) = 0\, ,
\]
which has the solution
\[
\varphi  \left( t \right) = A e^{ - 0.4t}  + B e^{ - 2.6 t} \, .
\]
For some real constants $A $, $B$. It is clear that in this approximation, the field $\varphi$ decreases in time until vanishes.

\vspace{\baselineskip}
However, $t$ is measured in unit of Plank mass, so $t=1$ is the time of value $m_{pl}^{-1}$ which is a large value, thus the  slow rolling period is long, but if a vacuum expectation value $\left\langle 0  \right|\varphi ^2 \left| 0  \right\rangle  = \varphi _0^2  \ne 0$ appears, the scaling symmetry breaks and a new Lagrangian (equation (\ref{eq:1})) takes place instead.

\section{Stability of ground state value of scalar field and energy}\footnote{This section is not included in the published edition.}
We have seen that there is a non-zero positive value of the energy density of the scalar field $\varphi$, this value $\rho _0 >0$ is given in the critical point $\dot \varphi=0$. But in order to relate $\varphi_0\ne0$ to quantum phenomena(i.e, vacuum expectation value), we need $\rho _0$ be stable and do not depend on time $\eta=\ln\left(a(t)\right)$. So we can regard $\varphi_0\ne0$ as a global constant value that can be given by $\varphi _0^2  = \left\langle \Omega  \right|\hat{ \varphi} ^2 \left| \Omega  \right\rangle  > 0$, for a ground state function $\left| \Omega  \right\rangle $. But we need to relate $\hat{ \varphi} $ and $\left| \Omega\right\rangle $ to a quantum phenomena which is global and does not depend on any geometry.

\vspace{\baselineskip}
From the charge equation (\ref{eq:4}) and constraint equation (\ref{eq:7}), we obtain at the critical point $\dot \varphi=0$, $\varphi=\varphi_0 \ne0$ the relations
\[
 \left. Q \right|_c  = Q =  - 12ka_0^3 H_0 \rho _0  \Rightarrow \rho _0  =  - \frac{Q}{{12ka_0^3 H_0 }}\,;\quad Q <0  \, ,
\]
and
 \begin{equation}\label{eq:a3} 
 \rho _0  - h_0 \rho _0  + \frac{{\rho _0^{(m)} }}{{a^3_0 }} = 0    \Rightarrow  \rho _0  = \frac{{\rho _0^{(m)} }}{{\left( {h_0  - 1} \right)a_0^3 }}\,  .
\end{equation}
These two equations imply
\[
 - \frac{Q}{{12kH_0 }} = \frac{{\rho _0^{(m)} }}{{\left( {h_0  - 1} \right)}} \, ,
\]
and by using $h=3k H^2$, we obtain
\begin{equation}\label{eq:a2}
H_0  = -\frac{{2\rho _0^{(m)} }}{Q} + \sqrt {\left( {\frac{{2\rho _0^{(m)} }}{Q}} \right)^2  + \frac{1}{{3k}}} >0\,;\quad  -Q>0 \, .
\end{equation}
It is clear that $H_0$ does not depend on the scale factor $a_0$, also it is global by its dependence only on the constants $k$, $Q$ and $\rho _0^{(m)}$ which are global by the meaning that they classify the solutions(do not depend on time).

\vspace{\baselineskip}
Therefore $H_0$ is global constant value. But in other side, we have 
\[
a\left( t \right) = a\left( 0 \right)e^{\int {H\left( t \right)dt} }\,  .
\]
Regarding the scaling symmetry, transformations (\ref{eq:18}), and before reaching the critical point $\dot \varphi=0$, $\varphi_0=\varphi(a_0)\ne 0$(in vicinity of it), we have the more general solution 
\[
a\left( t \right) = a\left( 0 \right)e^{2\alpha  + \int {H\left( t \right)dt} } \, ,
\]
for any real arbitrary constant $\alpha$. And according to equation (\ref{eq:19}), $h'\approx 0$ and so $H'\approx 0$ when $\rho^{(m)}_0/a^3_0\rho_0<<1$. Thus in vicinity of the critical point $\dot \varphi=0$, $\varphi_0=\varphi(a_0)\ne 0$, we use the value (\ref{eq:a2}) of $H_0$ to approximate $a\left( t \right) $ to
\[
a\left( t \right) = Ae^{2\alpha  + H_0 t} \, ,
\]
for some constant $A>0$. If we let the critical point $\dot \varphi=0$, $\varphi(a_0)=\varphi_0\ne 0$ be reached in time $t=t_0$, we obtain 
\[
a_0  = a\left( {t_0 } \right) = Ae^{2\alpha  + H_0 t_0 } \, .
\]
Now we can write
\[
2\alpha  + H_0 t_0=H_0 T_0
\]
and choose $\alpha$ such that $T_0=1$, by that we obtain
\[
a_0  = A e^{H_0 } \,  ,
\]
in the critical point $\dot \varphi=0$, $\varphi(a_0)=\varphi_0\ne 0$. But according to the equation (\ref{eq:a2}), $H_0=H(k, Q, \rho _0^{(m)} )$ which implies that $a_0 $ depends only on the globally constants $k$, $Q$ and $\rho _0^{(m)}$. Thus $a_0(k, Q, \rho _0^{(m)} )$ is also globally constant value and it also classifies the solutions, by that the energy density $\rho _0 $, equation (\ref{eq:a3}), depends only on the globally constants $k$, $Q$ and $\rho _0^{(m)}$, so it is also a globally constant value, not geometrical, thus it does not change under the universal expansion after passing the critical point $a=a_0$($\dot \varphi=0$). Therefore $\varphi _0^2  = \left\langle \Omega  \right|\hat{ \varphi} ^2 \left| \Omega  \right\rangle  $ and $\left| \Omega  \right\rangle $ are global structures, where $\rho _0 =\varphi _0^2 /2$.

\vspace{\baselineskip}
According to this discussion, we can think that $\rho _0$ is the vacuum expectation value of $\hat{ \varphi} ^2 $, where $\hat{ \varphi} $ is quantum field that does not depend on any geometry, as well as the quantum ground state $ \left| \Omega  \right\rangle $. By that the equality $\varphi _0^2/2 =\Lambda$(equations (\ref{eq:20})) is well defined and the cosmological constant $\Lambda$ in this view is global stable value, that it does not relate with the universal expansion, i.e, does not change under the universal expansion after passing the critical point $a=a_0$($\dot \varphi=0$).

\section{Summary and Conclusion}

In this paper, we have studied some novel aspects of cosmological dynamics of a quintessence scalar field non-minimally coupled to gravity in a spatially flat FRW background via the Noether Symmetry approach. We considered the non-minimal coupling between the scalar field and gravitational sector as $R L^{(\varphi)}$, that is essentially a subclass of the general Horndeski gravity and reduces to non-minimal derivative coupling in the case of kinetic dominance of the scalar field. We applied the Noether symmetry approach to the Lagrangian of the model and derived the corresponding Noether charge by exploring the status of the scaling symmetry in this framework. We adopted a suitable potential of the scalar field $\varphi$ and estimated the behaviour of the scale factor via scaling symmetry breaking in this setup. We treated the role of the Noether charge in the solutions of the scalar field and we have shown that by the universal positively accelerated expansion (especially an exponential expansion), the field $\varphi$ is always exponentially decreasing until reaching a critical point at $\dot \varphi=0$, that is, when $\varphi=\varphi_0 \ne0$, in which the global scaling symmetry breaks and the universal expansion is approximately in a constant rate $H=H_0$. Existence of scaling symmetry breaking violates the conservation of the corresponding charge, that is, $dQ/dt\ne0$ in the critical point $\dot \varphi=0$, $\varphi=\varphi_0 \ne0$. The existence of a non-vanishing constant positive value $\varphi_0$ at the critical point $\dot \varphi=0$ is necessary for fulfilling the constraint equation $\delta S /\delta N=0$. We have demonstrated that the critical point $\dot \varphi=0$, $\varphi=\varphi_0 \ne0$ is unique and stable in this setup and as an important result, we were able to relate the cosmological constant and gravitational constant via an identity, which is scaling symmetry breaking in the space $(a, \varphi)$. Finally we tried to show that the ground state energy density $\rho _0$ relates to quantum phenomena and globally stable.\\

{\bf Funding and/or Conflicts of interests/Competing interests:}

There is no funding and/or conflicts of interests/ompeting interests regarding this manuscript.\\

{\bf Data Availability Statement}:

 No Data associated in the manuscript.

\end{document}